\documentclass[aps,twocolumn,showpacs,preprintnumbers,prb,amsmath,amssymb,amsfonts,superscriptaddress,floatfix]{revtex4-1}

\usepackage[latin1]{inputenc}
\usepackage{graphics}
\usepackage{color}
\usepackage{amssymb}
\usepackage{amsmath}
\usepackage{overpic}
\usepackage{epstopdf}
\usepackage{bm}
\usepackage{graphicx}
\usepackage{subfigure}

\begin{document}

\title{Electronic and magnetic properties of perovskite
selenite and tellurite
compounds: CoSeO$_3$, NiSeO$_3$, CoTeO$_3$ and NiTeO$_3$}

\author{A. Rafi M. Iasir}
\thanks{These authors contributed equally}
\affiliation{Nuclear Engineering Program, University of Missouri,
Columbia MO 65211, USA}

\author{Todd Lombardi}
\thanks{These authors contributed equally}
\affiliation{Department of Physics and Astronomy, University of Missouri,
Columbia, MO 65211-7010, USA}

\author{Qiangsheng Lu}
\thanks{These authors contributed equally}
\affiliation{Department of Physics and Astronomy, University of Missouri,
Columbia, MO 65211-7010, USA}

\author{Amir M. Mofrad}
\thanks{These authors contributed equally}
\affiliation{Department of Biomedical, Biological and Chemical Engineering,
University of Missouri, Columbia MO 65211, USA}

\author{Mitchel Vaninger}
\thanks{These authors contributed equally}
\affiliation{Department of Physics and Astronomy, University of Missouri,
Columbia, MO 65211-7010, USA}

\author{Xiaoqian Zhang}
\thanks{These authors contributed equally}
\affiliation{Department of Physics and Astronomy, University of Missouri,
Columbia, MO 65211-7010, USA}
\affiliation{Jiangsu Provincial Key Laboratory of
Advanced Photonic and Electronic Materials,
Collaborative Innovation Center of Advanced Microstructures,
School of Electronic Science and Engineering,
Nanjing University, Nanjing 210093, China}

\author{David J. Singh}
\affiliation{Department of Physics and Astronomy, University of Missouri,
Columbia, MO 65211-7010, USA}
\affiliation{Department of Chemistry, University of Missouri,
Columbia, MO 65211, USA}

\email{singhdj@missouri.edu}

\date{\today}

\begin{abstract}
Selenium and tellurium are among the few elements that form $AB$O$_3$
perovskite structures with a four valent ion in the $A$ site. This
leads to highly distorted structures and unusual magnetic behavior.
Here we investigate the Co and Ni selenite and tellurite compounds,
CoSeO$_3$, CoTeO$_3$, NiSeO$_3$ and NiTeO$_3$ using first principles
calculations. We find an interplay of crystal field and Jahn-Teller
distortions that underpin the electronic and magnetic properties.
While all compounds are predicted to show an insulating G-type
antiferromagnetic ground state, there is a considerable difference
in the anisotropy of the exchange interactions between the Ni and
Co compounds. This is related to the Jahn-Teller distortion.
Finally, we observe that these four compounds show characteristics
generally associated with Mott insulators, even when described at 
the level of standard density functional theory. These are then dense bulk
band or Slater, Mott-type insulators.
\end{abstract}

\maketitle

\section{Introduction}

Perovskite oxides constitute an exceptionally broad class of compounds,
and exhibit many diverse properties and useful functionalities.
\cite{pena,lufaso}
This includes high temperature superconductors,
\cite{bednorz}
widely used ferroelectric and piezoelectric materials,
\cite{jaffe}
colossal magnetoresistive compounds, \cite{jonker,vonhelmolt}
ionic conductors, \cite{malavasi}
and many other important materials.
The structure type is characterized by an $AB$O$_3$ stoichiometry, based
on corner linked $B$O$_6$ octahedra arranged on a simple cubic lattice,
often distorted.
These distortions are key to the chemical flexibility of the 
perovskite structure. For example, $A$-site ions that are too
small for the site can be accommodated by coordinated tilting of the
corner linked $B$O$_6$ octahedra, thus allowing small ions to be inserted,
and at the same time providing a chemical tuning mechanism for modulating
the bond angles and properties that depend on them.
The chemical and structural tunability of this structure type
also leads to fascinating physics including quantum critical phenomena,
Mott and other metal insulator transitions, and magnetoelectric
materials.
The manganites in particular, but other materials as well,
have also highlighted the importance of coupling between structure
and electronic properties in perovskites. 
\cite{lufaso,jonker,pickett,kimura,khomskii}
In this regard, investigation of compounds with unusual perovskite chemistry
or structure is useful in understanding the range of possible behaviors.

Here we investigate the cobalt and nickel
containing selenite and tellurite compounds, CoSeO$_3$, CoTeO$_3$,
NiSeO$_3$ and NiTeO$_3$. These compounds, although known since
the 1970's,
\cite{kohn-74,kohn,kohn2}
have been relatively little studied.
These are highly unusual perovskites. In particular, they
have extremely strong distortions leading to very strong deviation
of the metal-O-metal bond angles from the ideal value of
180$^\circ$ and they have exclusively
divalent cations at the perovskite $B$-site, while normally
only partial occupancy of divalent $B$-site ions, counter-balanced by higher
valence ions on other $B$-sites can occur in perovskites. This is as
rationalized by the Pauling rules. \cite{pauling}
in which the corner sharing perovskite structure is stabilized by
repulsion between highly charged $B$-site ions.
Here, the unusual structures are presumably stabilized by covalent
interactions between O and the Se or Te, leading to a chemistry
intermediate between, on the one hand normal ionic perovskite crystals,
in this case stabilized by strong off-centering and lone pair
activity of the very small
Se$^{4+}$ and Te$^{4+}$ ions, and on the other hand salts based on
complex anions, \cite{kohn}
specifically, (SeO$_3$)$^{2-}$ or (TeO$_3$)$^{2-}$.

It is known that CoSeO$_3$, CoTeO$_3$, NiSeO$_3$ and NiTeO$_3$ all
have magnetic ground states. While transport
data and spectroscopy have not been reported,
based on their reported colors \cite{kohn}
they are probably insulating.
Unlike MnSeO$_3$, insulating behavior in these compounds is not
an obvious result from electron counting.
The corresponding Cu compounds show a strong dependence of magnetic
order on structure, including a cross-over from antiferromagnetic
to ferromagnetic behavior as the bond angle is distorted in the
CuTeO$_3$ -- CuSeO$_3$ alloy system. \cite{subramanian}

Remarkably, in the present study we find all four of these materials to
exhibit a novel type of insulating character, specifically they are predicted
to be band insulators at the PBE level both in their ground state and in
the disordered paramagnetic state. This is a characteristic
of Mott-type insulators, and so they may be called Slater Mott-type insulators.

\section{Methods and Structure}

The calculations presented here were done in the framework of density
functional theory (DFT) using the general potential linearized augmented
planewave (LAPW) method \cite{singh-book} as implemented in the WIEN2k code.
\cite{wien2k}
We used the generalized gradient approximation (GGA) of Perdew, Burke
and Ernzerhof (PBE).
\cite{pbe}
We tested different Brillouin zone samplings and did calculations with
different choices of LAPW sphere radii.
The main results shown here were obtained with LAPW sphere radii of
1.8 Bohr for Ni, Co, Se and Te, and
1.4 Bohr for O.
The LAPW plus local orbital basis sets were used with planewave sector
cutoffs, $K_{max}$ determined by the criterion $R_{min}$$K_{max}$=7,
where $R_{min}$ is the smallest, i.e. the O,
sphere radius. This leads to an effective
value of 9 for the metal, Se and Te atoms.

The four compounds studied are all reported to occur in an orthorhombic
$Pnma$ (spacegroup 62) structure with four formula units per unit cell.
This is a very common perovskite variant.
However, they are unusual in the size of the distortions from the
ideal cubic structure, which are extremely large in these materials.
Our calculations are based on the experimental lattice parameters.
The internal atomic coordinates in the unit cells were determined
by total energy minimization with a ferromagnetic ordering. We also
did calculations for antiferromagnetic orderings. However, the calculated
forces remained small independent of the specific ordering pattern, and so
we conclude that relaxation based on ferromagnetic ordering is sufficient.

\begin{figure}
\includegraphics[width=\columnwidth]{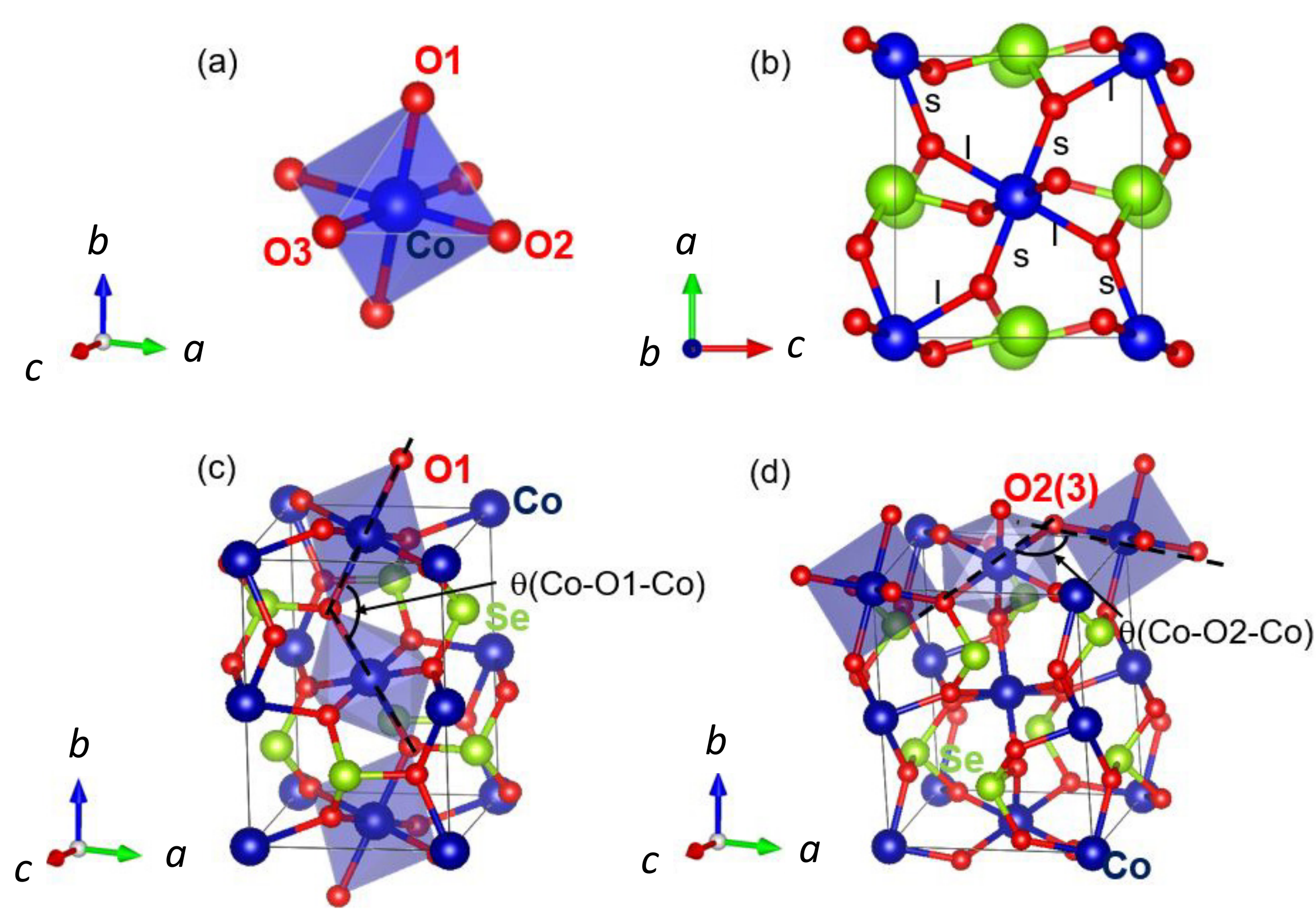}
\caption{\label{struct} Crystal structure of $Pnma$ orthorhombic CoSeO$_3$.
(a) CoO$_6$ octahedron, (b) view along $b$-axis, (c) arrangement of
octahedra in $ac$ plane and (d) view showing bond angles due to octahedral
rotation.}
\end{figure}

\begin{table}
\caption{Structural parameters of the four compounds, based on the
relaxed atomic positions (see text and Fig. \ref{struct}). $M$ is the metal
atom (Co or Ni).}
\begin{tabular}{lcccc}
\hline \hline
                    &  ~CoSeO$_3$~ & ~NiSeO$_3$~ & ~CoTeO$_3$~ & ~NiTeO$_3$~ \\
\hline
$M$-O1 (\AA)        & 2.129 & 2.110 & 2.079 & 2.072 \\
$M$-O2 (\AA)        & 2.211 & 2.146 & 2.287 & 2.189 \\
$M$-O3 (\AA)        & 2.056 & 2.058 & 2.082 & 2.101 \\
$\theta$($M$-O1-$M$) & 126.2$^\circ$ & 126.1$^\circ$ & 129.2$^\circ$ & 129.6$^\circ$ \\
$\theta$($M$-O2-$M$) & 131.3$^\circ$ & 132.0$^\circ$ & 133.7$^\circ$ & 134.6$^\circ$ \\
$\Delta_d$ & 0.00088 & 0.00030 & 0.00205 & 0.00055 \\
\hline
\end{tabular}
\label{tab-dist}
\end{table}

The structure of CoSeO$_3$, which is 
representative, is depicted in Fig. \ref{struct}.
The calculated structural data is summarized in Supplemental
Table S1.
As seen, the structures, while perovskite in nature with the characteristic
motif of corner sharing $M$O$_6$ octahedra,
are very strongly distorted from the cubic perovskite
structure. This is evident also in the reported experimental lattice
parameters, which deviate from the pseudo-cubic relationship,
$a$=$c$=$\sqrt{2}a^*$, $b$=2$a^*$=$\sqrt{2}a$, where $a^*$ is the effective
cubic lattice parameter.
The Co compounds have larger lattice parameters and bond lengths
than the corresponding Ni compounds.
This simply reflects the larger ionic radius of high spin Co$^{2+}$
compared to high spin Ni$^{2+}$. \cite{shannon}
Within the structure, all four $M$O$_6$ octahedra are equivalent, and
from the perspective of the octahedra and their connectivity,
the local structure is characterized by
three metal-O bond lengths and
two metal-O-metal bond angles, as illustrated in
Fig. \ref{struct}(a), \ref{struct}(c) and \ref{struct}(d).
Specifically, as illustrated, the three angles are Co-O1-Co, Co-O2-Co
and Co-O3-Co, but Co-O2-Co and Co-O3-Co are the same
(O1 is the connection
to the Co along the $b$-axis, while O2 and O3 are the O that
connect in the $ac$-plane directions).
These parameters are given in Table \ref{tab-dist}, along with
an octahedral distortion parameter, \cite{alonso-delta}

\begin{equation}
\Delta_d = {{1}\over{6}}\sum_{n=1,6} \left [ {{d_n - d}\over{d}} \right ]^2 ,
\end{equation}

\noindent where the $d_n$ are the six metal -- O distances in an octahedron,
and $d$ is the average.
This distortion parameter was previously used in the analysis of 
Jahn-Teller splittings in manganites. \cite{alonso-delta}
As seen, the bond angles characterizing the octahedral rotation are similar
for the four compounds, and are slightly lower
(corresponding to stronger rotation) for the selenite compounds,
consistent with the smaller size of Se$^{+4}$ compared to Te$^{+4}$.
The distortion of the octahedra on the other hand is sensitive to the metal
atom, and is substantially larger for the Co compounds as compared to the
corresponding Ni compounds.

\section{Electronic Structure}

\begin{figure}
\includegraphics[width=0.85\columnwidth]{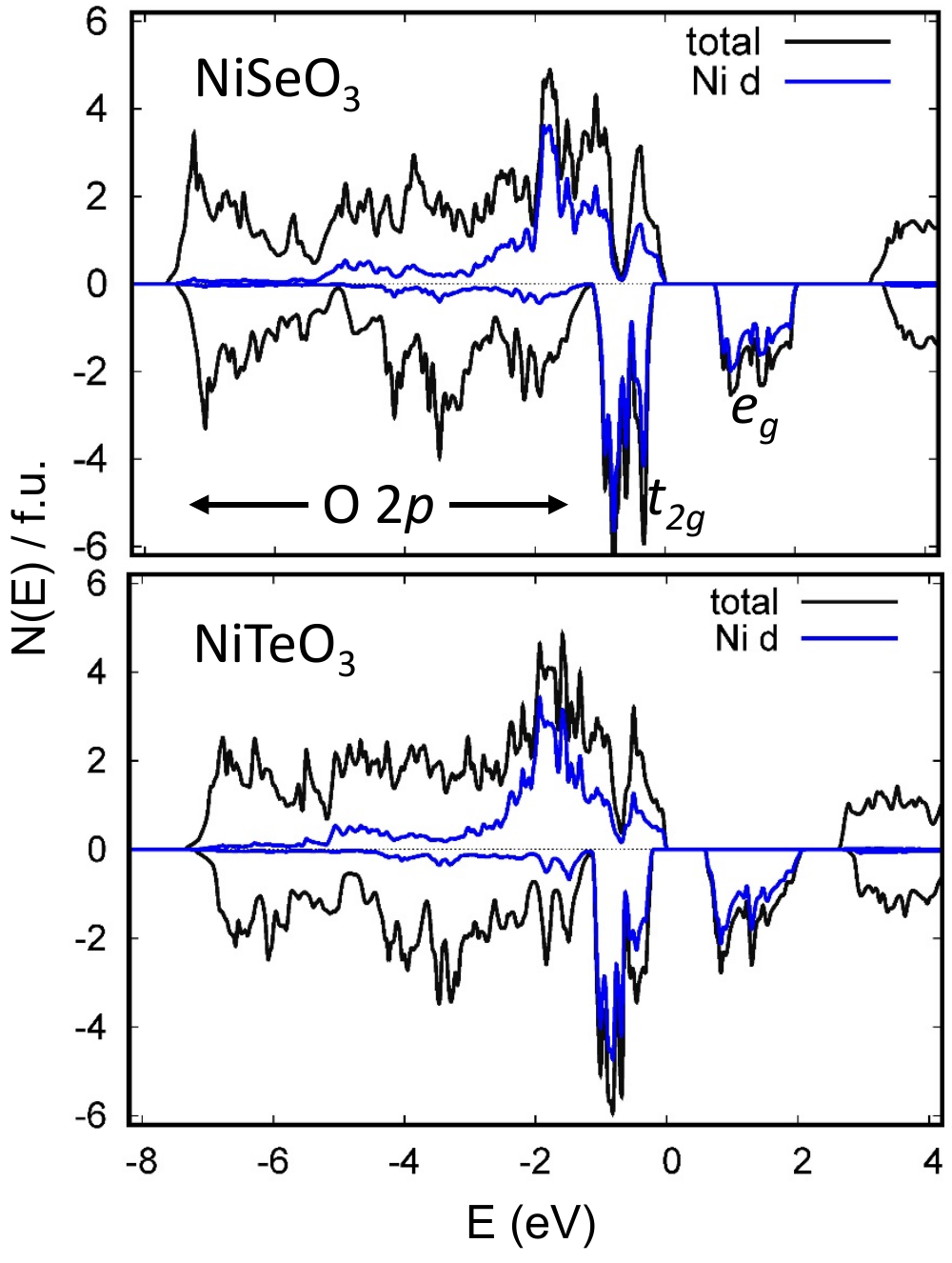}
\caption{\label{dos-f-ni} Electronic density of states
and Ni $d$ projections for non-ground
state ferromagnetic ordering for NiSeO$_3$ and NiTeO$_3$.
The projects are onto the LAPW spheres. The
energy zero is at the highest occupied state.}
\end{figure}

\begin{figure}
\includegraphics[width=0.85\columnwidth]{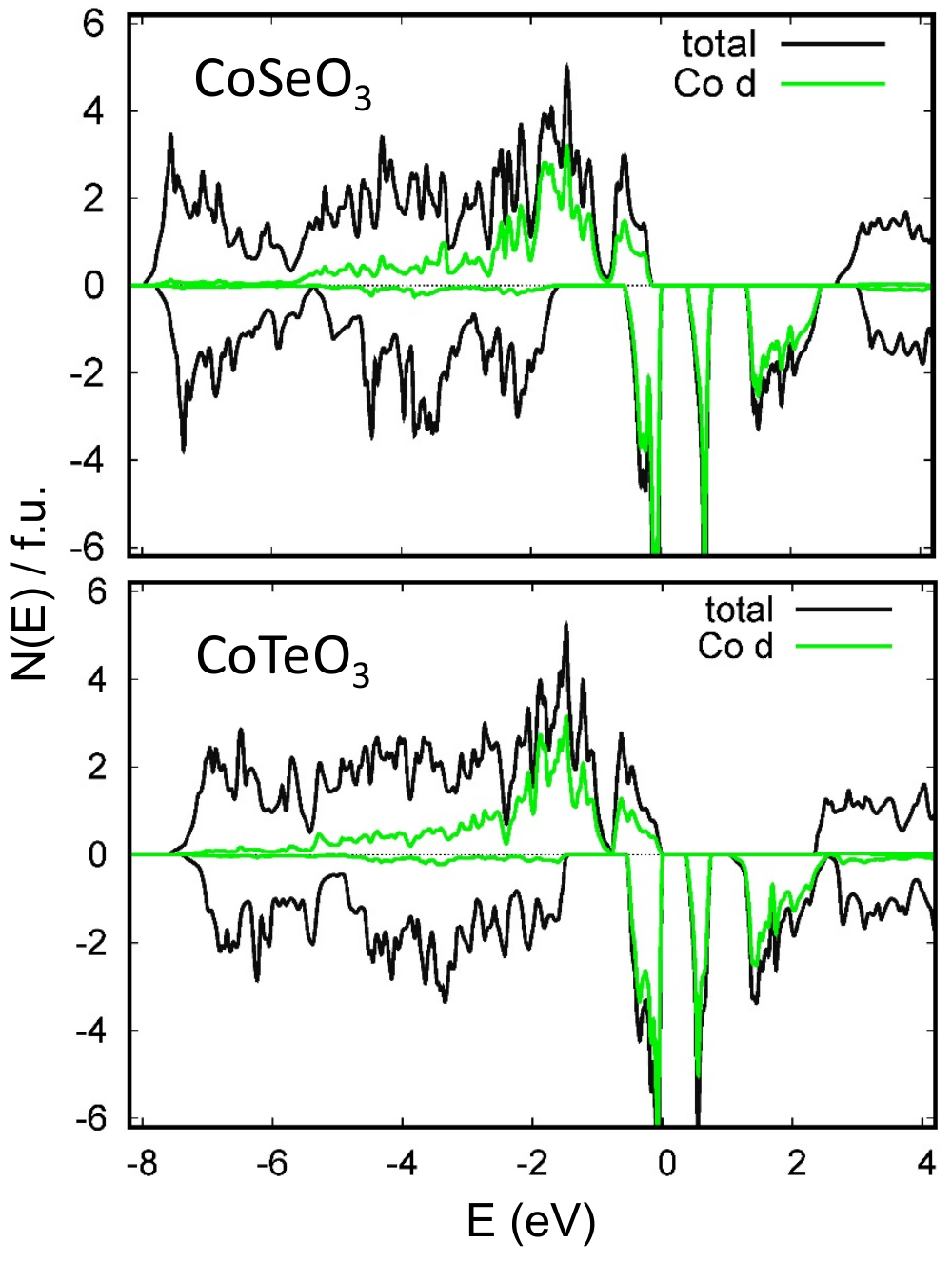}
\caption{\label{dos-f-co} Electronic density of states
and Co $d$ projections for non-ground
state ferromagnetic ordering for CoSeO$_3$ and CoTeO$_3$.}
\end{figure}

As mentioned, the structure relaxation was done using an assumed ferromagnetic
ordering and the standard PBE GGA density functional.
Ferromagnetic solutions were found for all four compounds
(note that as discussed below there are lower energy antiferromagnetic 
orderings).
The calculated electronic densities of states are shown along
with projections in Fig. \ref{dos-f-ni} and Fig. \ref{dos-f-co}.
Remarkably all the
compounds are insulating with this ferromagnetic ordering.
This is similar to what was found previously for MnSeO$_3$,
which has the $d^5$ ion Mn$^{2+}$,
forming a high spin antiferromagnetic insulating ground state.
\cite{michel,honer}
The spin magnetizations are 3 $\mu_B$ per formula unit for
CoSeO$_3$ and CoTeO$_3$ and 2 $\mu_B$ per formula unit for the
Ni compounds, NiSeO$_3$ and NiTeO$_3$.

The electronic structure of the four compounds in the valence energy range
consists of occupied O $2p$ derived bands from $\sim$ -7 eV to
$\sim$ -1 eV, with respect to the energy zero, which is set to
the highest occupied state. The chalcogen $s$ states are below the window shown.
The transition metal states show strong exchange splittings, as seen.
Additionally, there is a strong spin dependent hybridization between the
O $2p$ states and the transition metal $d$ states.
The spin dependence
is seen in the differences in shape of the majority and minority
$d$ character in Figs. \ref{dos-f-ni} and \ref{dos-f-co}.
It arises from the fact that the majority $d$ states overlap the top of the O $2p$
bands, while the minority $d$ states are mainly above the $2p$ bands.
However, while the minority spin shows weaker hybridization than the
majority, the hybridization is substantial in both spin channels.
This is evident from the sizable crystal field splitting of the
minority spin $d$ states, which amounts to $\sim$ 2 eV for both the
Co and Ni compounds
(note that in transition metal oxides crystal field splitting is
due metal-ligand hybridization).
The nominally unoccupied $p$ states of Se and Te
occur above the transition metal
bands starting at $\sim$3 eV for the Ni compounds and $\sim$2 eV for the Co
compounds.
These are formally the antibonding combinations of O $2p$ and Se/Te $p$
states from the (SeO$_3$)$^{2-}$ and (TeO$_3$)$^{2-}$
complex anions within a view of the crystal structure as a salt.

The insulating nature of the Ni compounds can be understood
in a standard crystal field scheme.
In an octahedral environment the 3$d$ states are crystal field
split into a lower lying $t_{2g}$ manifold with three states per spin
and a higher $e_g$ manifold with two states per spin.
The eight $3d$ electrons of Ni$^{2+}$ suffice to fill the
majority spin $d$ levels plus the minority spin $t_{2g}$ level.
This then can lead to insulating behavior provided that the
band widths are narrow enough to leave clean gaps between
the minority spin $t_{2g}$ and $e_g$ crystal field levels.
This is evidently the case in ferromagnetic
NiSeO$_3$ and NiTeO$_3$ as can be seen from Fig. \ref{dos-f-ni}.
It should be noted, however, that this is unusual, since
ferromagnetic ordering is particularly favorable for hopping and therefore
is the ordering that typically has the largest band widths in oxides.
For example, as pointed out by Slater, \cite{slater}
it is possible to have insulating gaps in antiferromagnetic systems
due to band structure effects.
This includes the prototypical Mott insulator, NiO. \cite{terakura}
However, the insulating character of NiO in this band structure point
of view depends crucially on the particular magnetic order.
In contrast, in the present case
the bands are exceptionally narrow leading to narrow crystal field split
levels.
Thus the crystal field splitting is not washed out,
even with ferromagnetic order.
This is a consequence of the highly distorted perovskite structure that
includes very strongly bent metal-O-metal bonds.

More remarkably, we find that both CoSeO$_3$ and CoTeO$_3$ have insulating gaps
with ferromagnetic order. Co$^{2+}$ has one less electron than Ni$^{2+}$.
This means that there are only two minority spin electrons.
As a result, there
is one hole in the $t_{2g}$ crystal field level.
Such partial filling of the $t_{2g}$ level should normally lead to a metallic
state especially with ferromagnetic order.
However, in the Co compounds, unlike the corresponding Ni compounds,
we find a clean
gap in the minority spin $t_{2g}$ levels at 2/3 filling.
This is a Jahn-Teller splitting that results from the lowering of local symmetry
due to the distortion of the CoO$_6$ octahedra. This corresponds to the
more distorted octahedra that we find in the relaxed structures of the Co
compounds, characterized by the parameter, $\Delta_d$ (Table \ref{tab-dist}).

We emphasize that here we performed standard GGA calculations and obtain these
insulating states. This differs from calculations done with additional
terms, such as in DFT+$U$ methods, where an interaction that increases
the separation of occupied and unoccupied $d$ levels is applied.
\cite{anisimov}
Furthermore, we use the relaxed atomic positions
obtained with standard GGA calculations,
and still obtain a sufficient Jahn-Teller
distortion to open a clean gap in the minority spin $t_{2g}$ levels of the
Co compounds, even for the ferromagnetic ordering.
This is particularly remarkable considering that the orbital involved is
Co $d$ - O $p$
$\pi$ antibonding $t_{2g}$,
which generally may be expected to be much less Jahn-Teller active than the
$\sigma$ antibonding $e_g$ orbitals.

\section{Magnetism}

We now discuss the magnetic properties. Experiment shows all
four compounds to be antiferromagnetic based on temperature dependent
susceptibility measurements.
\cite{kohn-74,kohn,kohn2}
However, the particular magnetic order has not been established.
The primitive unit cell contains four transition element atoms.
Within one unit cell it is therefore possible to consider four collinear
magnetic orders, denoted F, G, A and C,
and corresponding to ferromagnetic order,
nearest neighbor antiferromagnetic order,
ferromagnetic layers stacked antiferromagnetically along the
crystallographic $b$-axis,
and nearest neighbor antiferromagnetic layers,
stacked ferromagnetically along $b$
to yield ferromagnetic chains along $b$, respectively.
We calculated the total energy and electronic structures for all these orders,
for each compound. These energies are given relative to the ferromagnetic
order in Table \ref{tab-mag}.
The last row is the ratio of the experimental $k_BT_N$ to the
energy difference between the F and ground state G type orders.
This energy difference
is a measure of the average superexchange strength in a nearest
neighbor model.
In all cases the G-type order had the lowest energy, and all antiferromagnetic
orders had lower energy than the ferromagnetic order. Within a nearest
neighbor superexchange scheme, this shows that all the bonds of a
metal atom to its six nearest neighbor metal atoms have
antiferromagnetic coupling.

In all cases and for all orders studied we find insulating gaps.
Furthermore, in all cases, the fundamental band gap increased
in going from the ferromagnetic to the antiferromagnetic ground state.
Band gaps for the different orders are presented in the Supplemental Table S2.
The Se compounds have larger band gaps than the corresponding Te compounds.
Also the band gap increase with change of the order to antiferromagnetic
is much
stronger for the Ni compounds.
In these Ni compounds the gap is between the occupied minority
spin $t_{2g}$ or majority $e_g$ (for ferromagnetic)
manifolds and the unoccupied minority $e_g$
manifold and the increase amounts
to $\sim$0.5 eV.
In the Co compounds, the gap is between the lower and
upper Jahn-Teller split $t_{2g}$ sub-manifolds, and the
increase amounts to less than 0.1 eV.
This reflects the superexchange mechanism, which involves the $e_g$ levels
as discussed below.
We note that the above places the magnetism in these
compounds in the local moment (as opposed to the itinerant) limit.

\begin{table}
\caption{Magnetic energies $E_{mag}$,
in eV per formula unit, relative to the ferromagnetic ordering.
\#AF is the number of antiferromagnetic
nearest neighbors for each transition metal. The ratio on the
last line is the ratio of 
the experimental $k_BT_N$
\cite{kohn-74,kohn,kohn2}
to the energy difference between the F and G orderings.}
\begin{tabular}{cccccc}
\hline \hline
Order & \#AF  &  ~CoSeO$_3$~ & ~CoTeO$_3$~ & ~NiSeO$_3$~ & ~NiTeO$_3$~ \\
\hline
F & 0 & 0 & 0 & 0 & 0 \\
A & 2 & -0.0013 & -0.0064 & -0.0226 & -0.0240 \\
C & 4 & -0.0263 & -0.0300 & -0.0516 & -0.0615 \\
G & 6 & -0.0275 & -0.0337 & -0.0735 & -0.0821 \\
ratio & & 0.15 & 0.20 & 0.11 & 0.13 \\
\hline
\end{tabular}
\label{tab-mag}
\end{table}

As mentioned, all the compounds are antiferromagnetic, with a G-type
ground state.
Experimentally, \cite{kohn-74,kohn,kohn2} the Neel temperatures, $T_N$,
show considerable variation among these materials.
The reported values are 49 K, 78 K, 98 K and 125 K, for
CoSeO$_3$, CoTeO$_3$, NiSeO$_3$ and NiTeO$_3$, respectively.
It is of interest to develop understanding of these differences.

\begin{figure}
\includegraphics[width=\columnwidth]{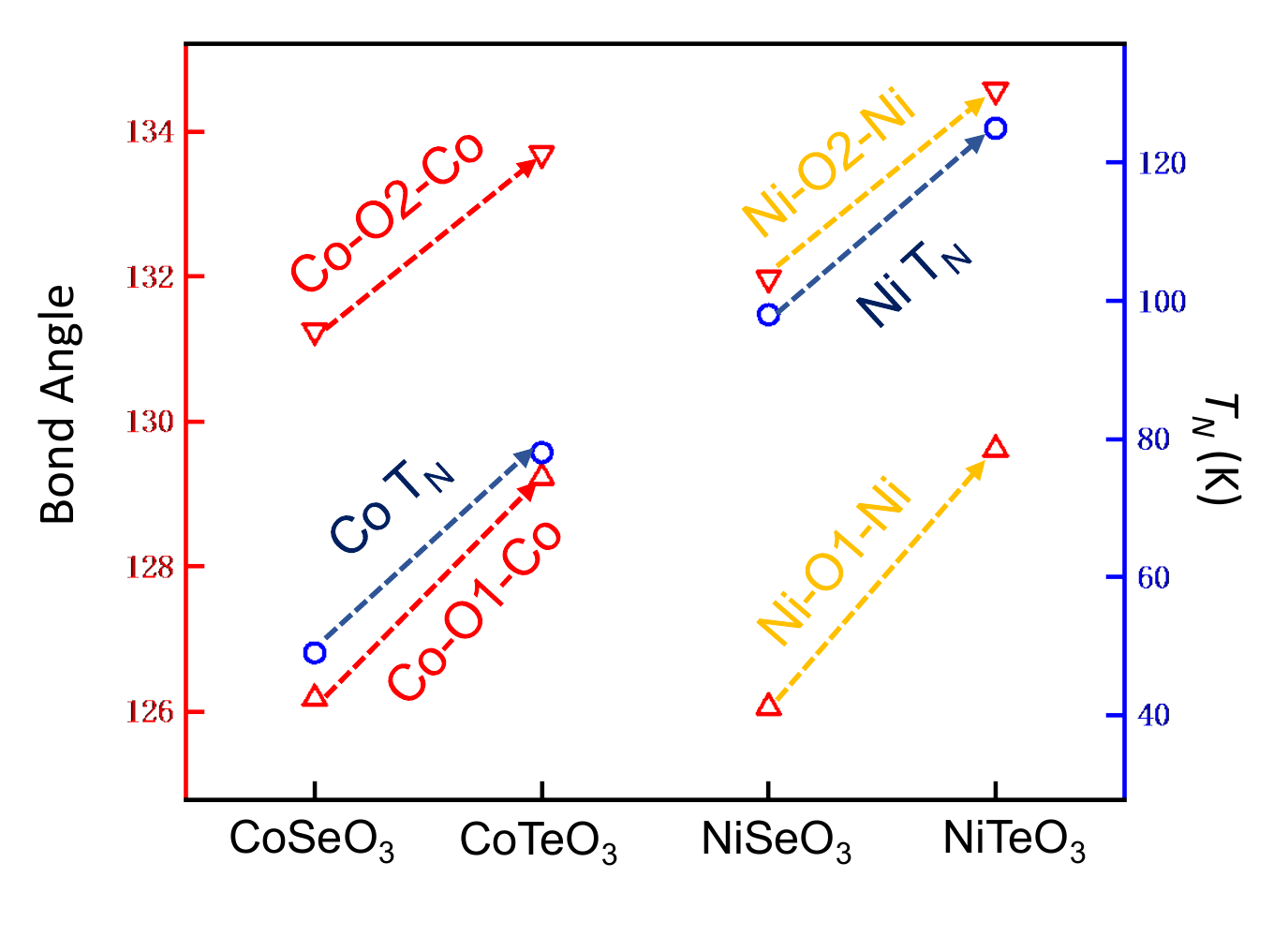}
\caption{\label{angle} Metal-O-metal bond angles from the
relaxed structure and experimental Neel temperature,
$T_N$, for the four compounds.}
\end{figure}

We begin with the structure.
Subramanian and co-workers emphasized the importance of the metal-O-metal
bond angles controlled by $A$-site size in explaining the cross-over from 
antiferromagnetism to ferromagnetism in the Cu(Se,Te)O$_3$ alloy system.
The Goodenough-Kanamori rules \cite{goodenough,kanamori,goodenough-book}
imply that the strongest antiferromagnetic
tendency for these materials should be for straight bonds.
In the present case, the perovskite framework is extremely distorted, with
bond angles very far from the ideal 180$^\circ$,
and so the validity of extrapolating from the properties of mildly
distorted perovskites may be questioned.
However,
the experimental $T_N$ are indeed higher for the Te compounds
than for the corresponding Se compounds. It is expected that the bonds will
be straighter for Te than Se due to the larger size of Te.
This is in fact the case based on our calculated structure data
(Table \ref{tab-dist}).
Fig. \ref{angle} shows the correlation between the two
different bond angles and
the experimental $T_N$ for the Co and Ni compounds.
This illustrates the trend, which is as expected from the Goodenough-Kanamori
rules. Furthermore, while the $T_N$ for the Ni compounds is consistently
higher than for the Co compounds, this is not a consequence of different
bond angles, e.g. from the smaller ionic radius of Ni, since in fact
the bond angles are similar. It is also of interest to observe that
the effect on the absolute $T_N$ of comparable changes in bond angle
are similar between the Co and Ni compounds.

Turning to the energies, it is notable that the energy difference
between the ferromagnetic and G-type antiferromagnetic ground state
follows the same order as the experimental values of $T_N$,
i.e. CoSeO$_3$ $<$ CoTeO$_3$ $<$ NiSeO$_3$ $<$ NiTeO$_3$.
However, there is not a simple proportionality, which might be expected
in an isotropic nearest neighbor Heisenberg model.
Instead, the Co compounds show higher $T_N$ than would be
expected from scaling down the Ni compound values according
to the F - G-type energy difference. This is seen in the
ratios given in the last line of Table \ref{tab-mag}.

This suggests a difference between the compounds, which
is somewhat surprising due to the fact that the basic features
of the electronic structure, especially as regards the $e_g$ orbitals
are similar between the Co and Ni compounds.
The $e_g$ states
are formally the $\sigma$ anti-bonding $d$ - O $p$ combinations.
\cite{goodenough,goodenough-book}
Therefore with fully occupied majority $e_g$ orbitals
and unoccupied minority $e_g$ orbitals, as in the
present compounds, the strongest superexchange channel
in perovskites involves the coupling of the $e_g$ states, which
as mentioned are similar between the four compounds.

One possible explanation could be in the chemical differences between Ni and
Co, for example different amounts of hybridization, and different
correlation strengths. One measure of the strength of hybridization
that is readily available is the crystal field splitting. We focus
on the minority spin in the ferromagnetic case, where the $t_{2g}$ and $e_g$
levels are separated from each other and from the O $2p$ bands.
We obtain the center of the corresponding peaks in the density of states
by integration of the first energy moment and thereby obtain the average
minority spin
crystal field splitting. In the Co compounds we include both sub-peaks of the
$t_{2g}$ bands in the calculation of the average $t_{2g}$ energy.
The obtained crystal field splittings are
1.72 eV, 1.69 eV, 2.00 eV and 1.97 eV, for
CoSeO$_3$, CoTeO$_3$, NiSeO$_3$ and NiTeO$_3$.
Thus the Ni compounds have larger crystal field splittings, indicating
stronger hybridization.
This may be the simple result of the fact that the Ni $d$ states are
closer in energy to the O $2p$ bands as seen in the
projected densities of states (Figs. \ref{dos-f-ni} and \ref{dos-f-co}).
In any case, it is consistent with stronger superexchange, since
superexchange depends on metal - ligand hybridization.
\cite{anderson}
This is as found in the present results.
This type of superexchange due to $e_g$ - $e_g$ interactions with occupied
$t_{2g}$ orbitals
increases the band gap for the antiferromagnetic state, since
antiferromagnetism both narrows the bands and shifts the $e_g$
level to higher energy.
In the case of the Co compounds, the gap is between sub-manifolds of
$t_{2g}$ states. In this case band narrowing can increase the gap, but
an up-shift of the $e_g$ level will not directly affect the gap,
explaining the smaller band gap increases upon antiferromagnetic ordering
in the Co compounds.

\begin{figure}
\includegraphics[width=0.85\columnwidth]{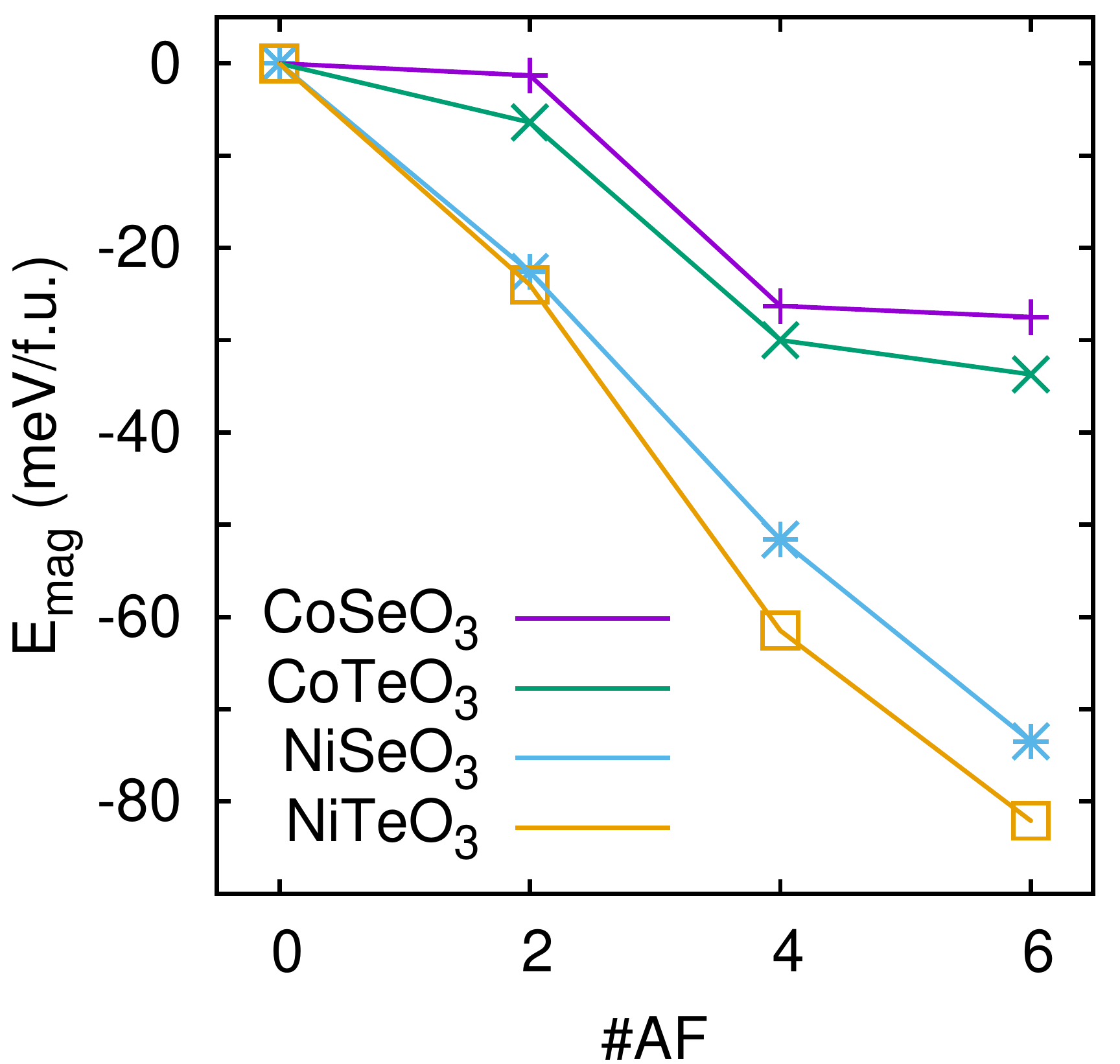}
\caption{\label{mag-e} Magnetic energy per
formula unit defined as the energy difference
from the ferromagnetic state for the F, A, C and G type orders,
which have 0, 2, 4 and 6 antiferromagnetic bonds per metal (\#AF),
respectively. Note the different behavior of the Co and Ni compounds.}
\end{figure}

This is related to the possible
explanation that strong correlations are present with
different values of $U$ that could be included in PBE+$U$
calculations with $U$ fit experiment.
However, while $U$ generally can be used to reduce exchange
couplings and fit $T_N$, and typically
some $U$ improves agreement with experiment in most
transition metal oxides, it would be desirable to have a more
first principles explanation if possible.
Also, the apparently stronger hybridization in the Ni compound as
indicated by the crystal field, does not support a view that the
Ni compound should have a larger $U$, since in general hybridization
provides screening, which reduces the effective $U$.
In any case, these results suggest that there may be a difference
in the magnetic behavior also at the PBE level.

As mentioned, the local transition metal environment is more
strongly distorted in the Co compounds due to the Jahn-Teller
effect.
This leads to anisotropy in the exchange interactions. This can
be seen in the pattern of energy differences in Table \ref{tab-mag}.
These energies are plotted in Fig. \ref{mag-e}.
The results show that the magnetic energy varies nearly linearly with
the number of antiferromagnetic bonds in the Ni compounds.
In contrast, antiferromagnetic bonds in the $ac$-plane strongly
lower the energy for the Co compounds. However in the Co compounds
the energy differences between the F and A order and between the C and G
orders are small. This means that the $b$-axis interactions are comparatively
quite weak.
Consequently, while all the
compounds have orthorhombic $Pnma$ symmetry,
NiSeO$_3$ and NiTeO$_3$ are nearly isotropic from the point of view of
their exchange interactions, but CoSeO$_3$ and CoTeO$_3$ are very
anisotropic, and are more like layered materials from the point of
view of exchange couplings.
Based on the obtained energetics, the interactions in-plane are similar
between the Ni and Co compounds 
(note especially the energy differences between A-type and C-type order,
which probe this).
Thus the Ni and Co compounds have very different anisotropy of the
exchange interactions, and so a simple scaling of $T_N$ with the F - G
energy difference is not expected between the Co and Ni compounds.

\section{Summary, Discussion and Conclusions}

To summarize the results, we find, using
standard density functional calculations at the PBE level,
that these compounds are
local moment antiferromagnetic insulators. The Ni compounds
are insulating because of the narrow bands and sizable crystal field
splitting. The Co compounds are also insulating due to a Jahn-Teller
distortion of the octahedra.
These calculations show near isotropic exchange interactions in the
Ni compounds, and a strong anisotropy with weak interactions along the
$b$-axis for the Co compounds. Remarkably, we find that these
compounds are predicted to be insulators regardless of the specific
magnetic order, even though we did standard PBE calculations.

A key characteristic of Mott insulators that distinguishes them
from Slater insulators is the fact that both the ordered magnetic
state and the higher temperature paramagnetic state are insulating
with similar band gaps.
This is a commonly applied experimental test for whether or not
a material is a band insulator (i.e. Slater insulator)
or a correlation driven Mott insulator.
As discussed below, the present compounds mix these two categories
as they satisfy the experimental test for a Mott insulator, but
can do so for purely band structure reasons.

In local moment systems, such as the compounds discussed here, the
phase transition is from an ordered antiferromagnetic state at low $T$
to a high temperature paramagnetic state where the moments persist
but are disordered. The spatial homogeneity of the paramagnetic
state may then be regarded as due to temporal variations of the
moments, that at any given time exist though without long range order.
If the temporal variations are not too fast, one may approximate this
state by thermodynamic averages over various disordered configurations.
This is the so-called disordered local moment picture, which
is very successful in describing both local moment materials and
even materials such as iron that have some itinerant character.
\cite{staunton,mizia,staunton-rev}

In the present selenite and tellurite
compounds, all configurations have band gaps, with approximately
similar values.
Therefore, a gap is expected for any disordered configuration and
therefore a disordered local moment picture will lead to gaps
in the paramagnetic states.
This includes the Co compounds where a Jahn-Teller distortion is present.

Thus these compounds present the unusual situation where one has local
moment magnetism with insulating character that persists above the
ordering temperature without the need for correlations beyond the standard
PBE level.
Recently, Zunger and co-workers, \cite{trimarchi,varignon,varignon-2}
have used a closely related concept to explain properties of transition metal
oxides, including monoxides and perovskites.
They used special quasi-random
structures to model the high temperature state. They did
calculations allowing for local
heterogeneity consistent with the quasi-random structures.
However, in the work it was necessary to either include a
correlation effect via a parameter $U$ or to use a density functional
such as SCAN that enhances the separation of occupied and unoccupied
$d$ levels, similar to a $U$ parameter, \cite{fu} leading to poor
descriptions of itinerant magnets such as iron. \cite{fu-fe}
Here we identify four bulk oxides, where the properties are characteristic
of a Mott insulator, even at the level of
standard PBE calculations, not including
any additional correlation terms.
This is not to say that the detailed agreement with experiments, when
these become available,
would not be improved by additional terms, such as $U$.
However, our results point out an interesting possibility, realized
in these compounds, that suggests further experimental
investigations of their detailed properties, 
especially band gaps and spectra, as well as spin excitations.
These may then be compared with theoretical results and used to
constrain models.
Spectroscopy in particular will be helpful in determining to what
extent the band or Slater Mott-type insulating picture developed above
is responsible for the properties of the materials, and conversely
how important the Coulomb interactions are in determining the spectra.

\acknowledgments

This paper results from work and analysis done by ARMI, TL, QL, AMM, MV and XZ,
in a project for
the course ``Structure, Electronic Structure and the Properties of Condensed
Matter" in the Department of Physics and Astronomy
at the University of Missouri, for which DJS was the instructor.
DJS is grateful for support from the U.S. Department of Energy,
Basic Energy Sciences, grant DE-SC0019114,
and for useful discussions with Alex Zunger.

\bibliography{project}

\end{document}